\documentclass[a4paper,twoside]{article}

\usepackage{epsfig}
\usepackage{calc}
\usepackage{amssymb}
\usepackage{amstext}
\usepackage{amsmath}
\usepackage{amsthm}
\usepackage{multicol}
\usepackage{pslatex}
\usepackage{algorithm2e}
\usepackage[bottom]{footmisc}
\usepackage{natbib}
\usepackage{graphicx}
\usepackage{caption,subcaption}
\usepackage{soul}
\usepackage{multirow}
\usepackage{hhline}
\usepackage{enumitem}

\usepackage{SCITEPRESS}
\usepackage[hidelinks]{hyperref}

\begin{document}

\title{Usability Evaluation and Improvement of a Tool for Self-Service Learning Analytics}

\author{
\authorname{Shoeb Joarder \orcidAuthor{0000-0003-4591-9742}, 
Mohamed Amine Chatti \orcidAuthor{0000-0002-1311-7852} 
and Louis Born}
\affiliation{Social Computing Group, Faculty of Computer Science, University of Duisburg-Essen, Forsthausweg 2, 47057 Duisburg, Germany.}
\email{\{shoeb.joarder, mohamed.chatti\}@uni-due.de,  louis.born@stud.uni-due.de}
}

% The paper must have at least one keyword. The text must be set to 9-point font size and without the use of bold or italic font style. For more than one keyword, please use a comma as a separator. Keywords must be titlecased.

\abstract{
Self-Service Learning Analytics (SSLA) tools aim to support educational stakeholders in creating learning analytics indicators without requiring technical expertise. 
While such tools promise user control and transparency, their effectiveness and adoption depend critically on usability aspects.
This paper presents a comprehensive usability evaluation and improvement of the \textit{Indicator Editor}, a no-code, exploratory SSLA tool that enables non-technical users to implement custom learning analytics indicators through a structured workflow.
Using an iterative evaluation approach, we conduct an exploratory qualitative user study, usability inspections of high-fidelity prototypes, and a workshop-based evaluation in an authentic educational setting with $n = 46$ students using standardized instruments, namely System Usability Scale (SUS), User Experience Questionnaire (UEQ), and Net Promoter Score (NPS).
Based on the evaluation findings, we derive concrete design implications that inform improvements in workflow guidance, feedback, and information presentation in the \textit{Indicator Editor}.
Furthermore, our evaluation provides practical insights for the design of usable SSLA tools.
}

\onecolumn \maketitle \normalsize \setcounter{footnote}{0} \vfill

%%%%%%%%%%%%%%%%%%%%%%%%%%%%%%%%%%%%%%%
\section{Introduction}\label{introduction}
%%%%%%%%%%%%%%%%%%%%%%%%%%%%%%%%%%%%%%%
Recent work in learning analytics (LA) has increasingly emphasized the importance of involving educational stakeholders (e.g., students, teachers, researchers, institutions) in the design and use of LA systems, giving rise to the emerging research area of human-centered learning analytics (HCLA) \citep{buckingham2019human,chatti2020design}. 
HCLA promotes the active participation of educational stakeholders across different stages of the LA lifecycle, including requirements elicitation, system development, evaluation, and post-deployment use \citep{buckingham2024human,chatti2021designing}. 
Existing HCLA studies report promising results from participatory and co-design approaches that incorporate stakeholder perspectives into LA system design. 
Nevertheless, the active involvement of end-users in the implementation of LA indicators remains largely unexplored \citep{alfredo2024human,topali2025designing,joarder2024anocode}. 
The LA indicator implementation task is often time-consuming and resource-intensive, frequently requiring specialized technical skills and infrastructure \citep{buckingham2024human}.
To address this issue, \citet{joarder2025human} introduced the concept of Self-Service Learning Analytics (SSLA), which aims to support lay educational stakeholders without programming experience in implementing LA indicators in an easy manner. 
The authors further presented the \textit{Indicator Editor}, a no-code, interactive SSLA tool that enables students and teachers with data analysis and visualization knowledge to implement LA indicators with minimal effort by providing access to the data and the analysis and visualization methods underpinning the indicators. 
The \textit{Indicator Editor} was then evaluated through a small qualitative user study to investigate its effects on users' perceptions of control, transparency, trust, satisfaction, and acceptance. 
However, the study did not cover empirical evaluations that systematically examine usability and user experience (UX) aspects during an LA indicator implementation exercise facilitated by the \textit{Indicator Editor}. 
While the \textit{Indicator Editor} supports flexibility and user agency, it also introduces complex interaction tasks, including selecting data sources, configuring analyses, and interpreting visualizations. 
Therefore, a usability investigation of the tool is also needed. To achieve this, in this work, we present a comprehensive usability and UX evaluation of the \textit{Indicator Editor}. 
Using an iterative evaluation approach that combines qualitative user studies, usability inspections, and a workshop-based evaluation in an authentic educational setting with standardized instruments, namely System Usability Scale (SUS) \citep{brooke2013sus}, User Experience Questionnaire (UEQ) \citep{laugwitz2008construction,schrepp2014applying}, and Net Promoter Score (NPS) \citep{keiningham2008holistic}, we provide empirical insights into usability challenges and derive design implications for improving the \textit{Indicator Editor} into its final user interface.

%%%%%%%%%%%%%%%%%%%%%%%%%%%%%%%%%%%%%%%
\section{Related Work and Context}\label{background}
%%%%%%%%%%%%%%%%%%%%%%%%%%%%%%%%%%%%%%%
The \textit{Indicator Editor} supports SSLA by enabling users to control the implementation of LA indicators by giving them access to the
data and the analysis and visualization methods underpinning the indicators \citep{joarder2025human}. 
It provides an interactive user interface that guides users through the full indicator implementation workflow. 
The workflow follows a progressive, step-by-step process that helps users work toward their goals by formulating questions and defining indicators to address them. The \textit{Indicator Editor} was developed through multiple iterations, applying Human-Centered Design (HCD) and usability design principles proposed by \citet{norman2013design}. 
In the following, we briefly summarize the functionalities of the \textit{Original User Interface of the Indicator Editor} presented in \citet{joarder2025human}.  
Figure \ref{fig:iterations-ui}a shows the main workflow components of this interface:
\begin{itemize}
    \item \textit{Dataset}: 
    Users define a data source by selecting a platform, an activity type, and a corresponding action via dependent multi-select dropdowns (Figure \ref{fig:iterations-ui}$a1$). 
    This selection determines the dataset used in all subsequent steps.
    \item \textit{Filters}: 
    Users refine the dataset using activity-, time-, and user-based filters organized in tabbed views (Figure \ref{fig:iterations-ui}$a2$). 
    These filters constrain the data to subsets relevant for analysis.
    \item \textit{Analysis}: 
    Users select an analysis method and map the filtered dataset to the required inputs and parameters of the method (Figure \ref{fig:iterations-ui}$a3$).
    \item \textit{Visualization}: 
    Users select a visualization library and chart type, and map analysis outputs to visualization inputs (Figure \ref{fig:iterations-ui}$a4$).
    \item \textit{Preview and Finalize}: 
    After previewing the result, users finalize and save the indicator (Figure \ref{fig:iterations-ui}$a5$). 
    Saved indicators can be revisited and previewed from the \textit{My Indicators} dashboard.
\end{itemize}
The \textit{Indicator Editor} has been evaluated with respect to transparency, trust, satisfaction, and acceptance \citep{joarder2025human}. 
However, usability and user experience (UX) have not yet been systematically assessed. In this work, we focus on the usability evaluation and improvement of the \textit{Original User Interface of the Indicator Editor} presented in \citet{joarder2025human}. To this end, we adopt an iterative, user-centered methodology to identify usability issues and iteratively improve the \textit{Indicator Editor}. 
Following an initial exploratory study, multiple prototypes were designed, evaluated, and refined, ultimately resulting in the \textit{Final User Interface of the Indicator Editor} (Figure \ref{fig:final-user-interface}). 
After each iteration, participant feedback was systematically incorporated into subsequent versions and re-evaluated. 
An overview of this iterative process, including the evolution of the main user interface components, is shown in Figure \ref{fig:iterations-ui}.
%%%%%%%%%%%%%%%%%%%%%%%%%%%%%%%%%%%%%%%%%%
\section{Qualitative User Study} \label{qualitativ-user-study}
\begin{figure*}[!ht]
	\centering
	\includegraphics[width=\linewidth]{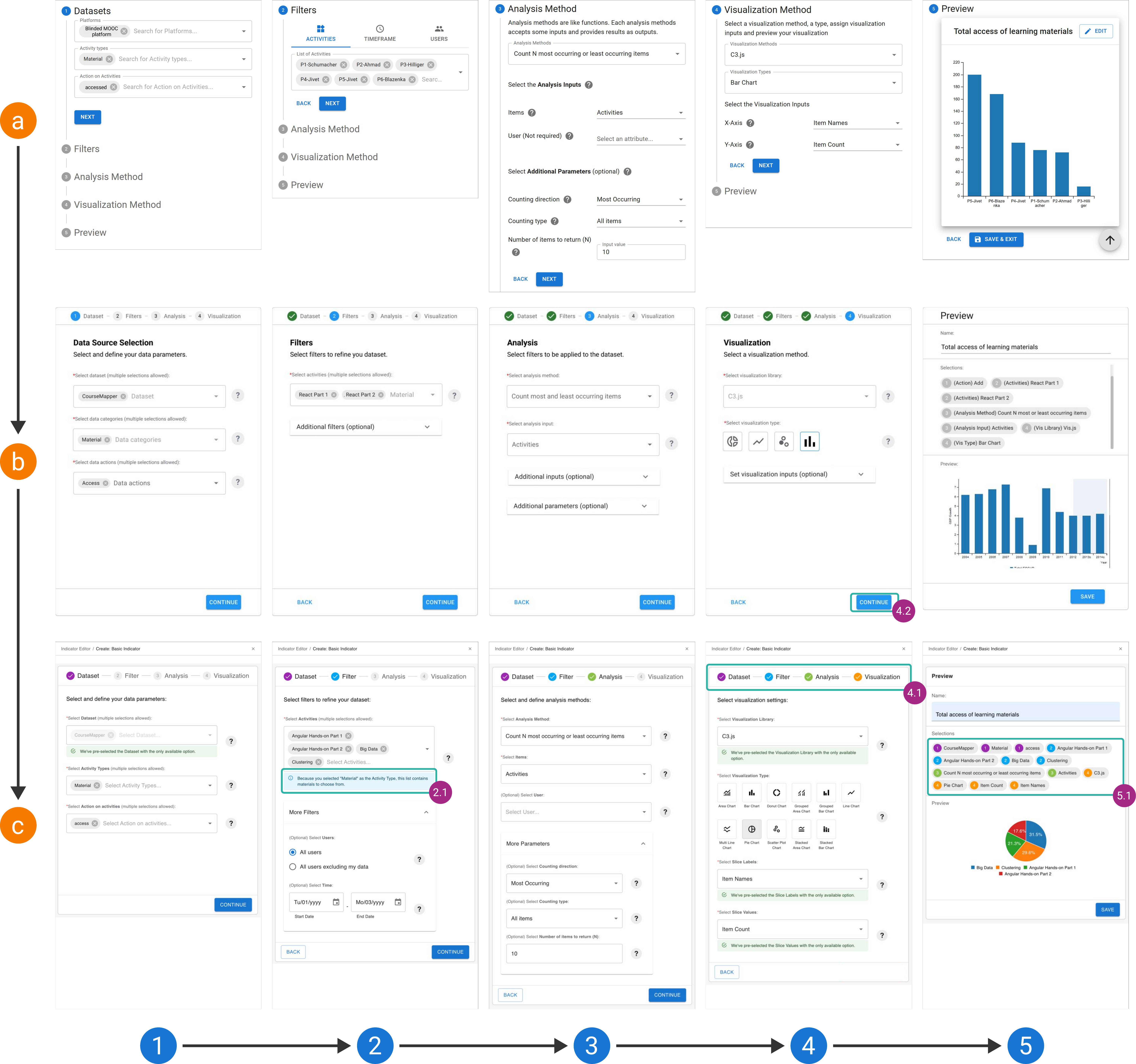}\\
    \caption{Evolution of the Indicator Editor interface across iterative design stages:
    \textbf{a}) the Original User Interface of the Indicator Editor, 
    \textbf{b}) the Initial High-Fidelity Prototype, and 
    \textbf{c}) the Intermediate High-Fidelity Prototype, developed in response to the identified usability findings.
    Across all iterations, the indicator creation workflow follows five main steps: 
    \textbf{1}) choose dataset, 
    \textbf{2}) apply filters, 
    \textbf{3}) select an analysis method, 
    \textbf{4}) select a visualization, and 
    \textbf{5}) preview and finalize the indicator.}
	\label{fig:iterations-ui}
\end{figure*}

%%%%%%%%%%%%%%%%%%%%%%%%%%%%%%%%%%%%%%%%%%
We started by conducting an exploratory qualitative user study to gain an in-depth understanding of potential end-users, their usage patterns, and their attitudes toward the \textit{Indicator Editor}. 
The study was based on the \textit{Original User Interface of the Indicator Editor} (Figure \ref{fig:iterations-ui}a), which had been deployed for testing purposes.
Semi-structured interviews were conducted remotely via Zoom, with the tool hosted on a cloud-based platform.
%%%%%%%%%%%%%%%%%%%%%%%%%%%%%%%%%%%%%%%%%

%%%%%%%%%%%%%%%%%%%%%%%%%%%%%%%%%%%%%%%%%
\subsection{Participants and Procedure}
Participants were recruited from individuals currently or previously enrolled at the local university who could participate independently and provided informed consent.
The study involved five participants: four master’s students (\textbf{ST}; $1$F, $3$M) and one teaching assistant (\textbf{TA}; $1$F), all with a background in computer science.

Participants first completed a short online questionnaire to collect demographic information and assess familiarity with data analytics, LA, and information visualization.
They were then introduced to the goals and core concepts of the \textit{Indicator Editor} and performed a set of representative tasks (dataset selection, filtering, analysis, and visualization) using a \textit{concurrent think-aloud protocol} \citep{ericsson2017protocol}.
This method enabled the capture of users' expectations, decision-making processes, and encountered difficulties during interaction \citep{fan2019concurrent,nielsen1994usability}.
After task completion, semi-structured interviews were conducted.
Sessions lasted approximately $30$–$45$ minutes, were screen-shared, and recorded with participants’ consent.

\subsection{Data Analysis}
The think-aloud sessions and interview recordings were transcribed verbatim following the transcription guidelines proposed by \citet{dresing2007qualitative}. 
To systematically analyze the qualitative data, we applied an inductive thematic analysis approach, as proposed by \citet{braun2006using}. 
The transcripts were iteratively coded using qualitative content analysis to identify recurring patterns related to users' expectations, interactions, reactions, and encountered usability issues. 
Coding and analysis were supported by the MAXQDA\footnote{https://www.maxqda.com} software. 
To ensure reliability, the coding scheme was validated through intracoder agreement following the procedure described by \citet{kuckartz2007einfuhrung}. 
The same researcher revisited the coding system after a time interval, reviewed all assignments, and resolved any inconsistencies. 
The resulting codes were subsequently grouped into higher-level themes, referred to as Findings ($F$), which informed the design decisions in later iterations.

\subsection{Results}
The analysis resulted in four central findings ($F$) that capture the main usability challenges experienced by participants. 
To support and illustrate these findings, representative participant quotes are included. 
All quotes are anonymized and referenced using role-based identifiers: \textbf{ST\#} denotes individual student participants, and \textbf{TA} refers to the teaching assistant.
\begin{enumerate}[label=\textbf{F\arabic*.}]
    \item \textbf{Clarity and feedback.}  
    Participants frequently expressed uncertainty about whether actions had been completed successfully and about their current position within the indicator creation workflow. 
    Missing or insufficient system feedback led to confusion, unnecessary backtracking, and reduced confidence in the interface. 
    For example, \textbf{ST1} remarked, \textit{``I did not see this [warning message box]''}, while the \textbf{TA} noted, \textit{``That [error feedback] was not very helpful. So I did not conclude from that I should select another type of visualization.''}
    Participants also emphasized the lack of interaction feedback, with the \textbf{TA} stating, \textit{``I miss the fact that you get feedback on what you click on.''}
    These statements highlight the need for clear, immediate feedback and explicit cues about system state and workflow progress.

    \item \textbf{Understandability through simpler language and in-context support.}  
    Technical terminology and unfamiliar concepts were perceived as significant barriers to understanding the system, particularly for users with limited technical expertise. 
    Participants reported that unclear labels and complex wording increased cognitive effort and slowed task completion. 
    As \textbf{ST2} explained, \textit{``In some places, it was not entirely clear to me what is meant by the terms.''}
    Similarly, \textbf{ST1} questioned ambiguous phrasing, stating, \textit{``But user not required [supposed to be an optional analytics method input], what do you mean here by this?''}, and \textbf{ST2} added, \textit{``The only issue was that I was unfamiliar with the terminology.''}
    These comments underline the importance of using plain language supported by contextual explanations, such as tooltips or embedded guidance, directly within the workflow.

    \item \textbf{Information overload reduction.}  
    Participants reported feeling overwhelmed by dense screens containing many options and data elements. 
    This information overload made it difficult to understand the indicator creation process and to make informed decisions. 
    For instance, \textbf{ST3} expressed confusion when encountering misleading or excessive information, stating, \textit{``Should I change the time frame? What is the time frame here?.''}
    Such experiences suggest that complex tasks should be decomposed into smaller, more manageable steps, with information revealed progressively to reduce cognitive load.

    \item \textbf{Visual, interaction, and information consistency.}  
    Inconsistencies in layout, interaction behavior, and visual elements disrupted participants' interaction flow and reduced predictability. 
    These inconsistencies forced users to relearn interaction patterns across different parts of the interface. 
    The \textbf{TA} highlighted this issue by noting, \textit{``Because when I selected the first one [Activity filter], it moved to here, but when I selected the second one [User filter], it stayed on top.''}
    Participants emphasized that a coherent, consistent visual and interaction design would make the system easier to understand and use, supporting more intuitive, efficient interactions.
\end{enumerate}
%%%%%%%%%%%%%%%%%%%%%%%%%%%%%%%%%%%%%%%%%%
\section{Usability Inspections} \label{usabilty-inspections}
%%%%%%%%%%%%%%%%%%%%%%%%%%%%%%%%%%%%%%%%%%
Building on the findings from the qualitative user study, we designed an \textit{Initial High-Fidelity Prototype} of the \textit{Indicator Editor} using the Figma
prototyping tool, as shown in Figure \ref{fig:iterations-ui}$b$. 
The goal of this iteration was to address the identified usability issues and to evaluate the effectiveness of the proposed design improvements before implementation.

\subsection{Participants and Procedure}
The prototype was evaluated with five new participants (3M, 2F), comprising three master's students and two teachers. 
Most participants were between 25 and 34 years old, with one participant in the 18–24 age range. 
All participants had an educational background in computer science or computer engineering and reported high familiarity with data analytics, LA, and information visualization.
The usability inspection sessions were conducted remotely via Microsoft Teams. 
Participants were asked to interact with the prototype while performing a set of representative tasks aligned with the indicator creation workflow. 
Sessions were moderated and recorded with participants' consent. 
Participants were encouraged to comment on their understanding of the interface, the clarity of system feedback, and any difficulties encountered during task execution.

\subsection{Results and Design Iterations}
Overall, the \textit{Initial High-Fidelity Prototype} received positive feedback, particularly for the improved workflow structure. 
Participants appreciated the persistent overview of their selections and the preview panel positioned to the right of the interface, which supported orientation and decision-making. 
The consistent placement of navigation buttons and the inclusion of help icons providing contextual explanations were also perceived as beneficial.

Despite these improvements, several usability issues were identified. 
One prominent issue concerned the selection of activities in the \textit{Filters} section. 
In the prototype, the available \textit{Activities} were derived from a combination of the selected \textit{Datasets}, \textit{Activity Types}, and \textit{Actions on Activities} specified in the \textit{Dataset} section. 
Participants reported confusion about how this list was generated and which selections influenced it. 
To address this issue, context-sensitive feedback messages were introduced in the subsequent iteration to clarify the origin of the activity list (\textbf{F1}, \textbf{F2}) (Figure \ref{fig:iterations-ui}$c2.1$).

Another recurring issue concerns the system's handling of changes made earlier in the process. 
Participants noted that even minor modifications to the \textit{Dataset} selection caused previously selected \textit{Activities} in the \textit{Filters} section to be automatically deselected without explicit notification, which was perceived as frustrating and unpredictable. 
To mitigate this problem, a pop-up feedback message was added to warn users when changing dataset options would affect existing activity selections, thereby making the consequences of such actions explicit (\textbf{F1}).

Participants also reported difficulties in locating and interpreting information presented in the \textit{Selections} panel. 
Although this panel displayed the choices made at each step of the workflow, users found the information insufficiently structured, making it difficult to quickly identify selections associated with specific steps. 
As a design response, selections were visually grouped and color-coded by workflow step in the \textit{Intermediate High-Fidelity Prototype} (Figures \ref{fig:iterations-ui}$c4.1$ and \ref{fig:iterations-ui}$c5.1$), improving scannability and consistency (\textbf{F3}, \textbf{F4}).

Finally, issues with manual regeneration of visualizations were identified. 
In the initial prototype, changes made in earlier steps were only reflected in the preview after users explicitly clicked the \textit{Continue} button (Figure \ref{fig:iterations-ui}$b4.2$). 
Participants occasionally forgot to regenerate the preview, leading to confusion and incorrect interpretations of the current configuration. 
In response, the \textit{Intermediate High-Fidelity Prototype} was designed to automatically update the preview whenever relevant parameters were changed, reducing interaction overhead and improving predictability (\textbf{F4}).

The design improvements identified during the usability inspections were subsequently implemented in the \textit{Intermediate High-Fidelity Prototype}, shown in Figure \ref{fig:iterations-ui}$c$. 
This version was developed using JavaScript and CSS frameworks, specifically React.js and Material UI.

%%%%%%%%%%%%%%%%%%%%%%%%%%%%%%%%%%%%%%%%%%
\section{Usability and User Experience Evaluation} \label{evaluation}
%%%%%%%%%%%%%%%%%%%%%%%%%%%%%%%%%%%%%%%%%%
We conducted a workshop to explore students' perceptions, expectations, and attitudes toward the \textit{Intermediate High-Fidelity Prototype} of the \textit{Indicator Editor} (Figure \ref{fig:iterations-ui}$c$), focusing on usability and user experience (UX).
%%%%%%%%%%%%%%%%%%%%%%%%%%%%%%%%%%%%%%%%%%
\subsection{Participants and Study Design}
%%%%%%%%%%%%%%%%%%%%%%%%%%%%%%%%%%%%%%%%%%
The evaluation was conducted within a university lab project involving bachelor's and master's students with at least basic knowledge of data analytics and information visualization. 
In total, $n = 46$ students ($26$F, $20$M), aged $18$--$34$ years, participated in the workshop.
Participants' study backgrounds were Computer Engineering ISE ($43\%$), Angewandte Kognitions- und Medienwissenschaft (Applied Cognitive and Media Studies) ($37\%$), Angewandte Informatik (Applied Computer Science) ($15\%$), and Wirtschaftsinformatik (Business Informatics) ($5\%$). 
Regarding the highest completed degree, $45\%$ of participants reported a bachelor's degree and $55\%$ a master's degree. 
All participants demonstrated sufficient English proficiency to complete the workshop tasks and questionnaires.
To contextualize the results, we assessed self-reported familiarity with data analytics, LA, and information visualization on a $4$-point scale ranging from $1$ (\textit{not so familiar}) to $4$ (\textit{extremely familiar}).
Responses of $1$--$2$ were grouped as \textit{not familiar}, and $3$--$4$ as \textit{familiar}. 
Based on this measure, $39\%$ of participants were familiar with data analytics, $22\%$ with LA, and $26\%$ with information visualization.

The evaluation was embedded within a university lab project and conducted as a hands-on workshop to collect structured feedback on the usability of the \textit{Indicator Editor}.
Participants first discussed LA concepts and potential indicators in the context of a CourseMapper\footnote{https://coursemapper.de}, and then implemented selected indicators using the \textit{Indicator Editor} that could better support their learning activities.
This ensured that the evaluated indicators were grounded in participants' own analytical goals rather than predefined examples.
The workshop lasted approximately 90 minutes.
After completing the hands-on tasks, participants filled out an online questionnaire including the System Usability Scale (SUS) \citep{brooke2013sus}, the User Experience Questionnaire (UEQ) \citep{laugwitz2008construction,schrepp2014applying}, and the Net Promoter Score (NPS) \citep{keiningham2008holistic}.
The SUS was used to assess perceived usability. 
The UEQ was selected to capture attractiveness and a broader range of pragmatic and hedonic aspects of UX. 
The NPS was included as a complementary, attitude-based measure to assess overall satisfaction and participants' willingness to recommend the \textit{Indicator Editor} to fellow students. 
Furthermore, participants answered three open-ended questions about positive aspects, limitations, and improvement suggestions:
\begin{enumerate}
    \item \textit{What do you like the most about the current state of the Indicator Editor?} 
    \item \textit{What do you like the least about the current state of the Indicator Editor?}  
    \item \textit{Do you have any suggestions to improve the tool?}
\end{enumerate}
All participants provided informed consent prior to participation.
%%%%%%%%%%%%%%%%%%%%%%%%%%%%%%%%%%%%%%%%%%
\subsection{Analysis and Results}
%%%%%%%%%%%%%%%%%%%%%%%%%%%%%%%%%%%%%%%%%%
The collected quantitative and qualitative data were analyzed to assess the usability of the \textit{Indicator Editor} from complementary perspectives. 
Quantitative results were derived from SUS, UEQ, and NPS, providing insights into perceived usability, attractiveness, pragmatic and hedonic quality, and users' willingness to recommend the system. 
These results were analyzed descriptively using established interpretation frameworks and benchmark guidelines. 
%%%%%%%%%%%%%%%%%%%%%%%%%%%%%%%%%%%%%%%%%%
\subsubsection{System Usability Scale (SUS)}
%%%%%%%%%%%%%%%%%%%%%%%%%%%%%%%%%%%%%%%%%%
The perceived usability of the \textit{Indicator Editor} was assessed using SUS. 
Overall, the system achieved a mean SUS score of $76.8$, which corresponds to a \textit{good} level of usability according to the interpretation framework \citep{brooke2013sus,bangor2009determining}. 
This result indicates that, on average, participants perceived the system as usable and were able to complete tasks successfully.
Individual SUS scores ranged from a minimum of $62.5$ to a maximum of $95.0$, reflecting variability in user perceptions. 
A subset of participants ($17.2\%$) assigned a SUS score of $85$ or higher, which is commonly interpreted as indicating excellent usability. 
In contrast, $13.8\%$ of participants reported SUS scores below $70$, suggesting that a small group experienced notable usability challenges when interacting with the system.
Although individual SUS items are not intended to be interpreted in isolation, the overall score can be contextualized using established grading scales. 
Based on the adjective ratings and grade equivalents reported by \citet{bangor2009determining}, the obtained mean score corresponds approximately to a grade $C$, indicating an acceptable but improvable level of usability. 
Taken together, the SUS results suggest that while the \textit{Indicator Editor} provides a generally usable interaction experience, certain aspects of the interface still hinder usability for a subset of lower-scoring users.
%%%%%%%%%%%%%%%%%%%%%%%%%%%%%%%%%%%%%%%%%%
\subsubsection{User Experience Questionnaire (UEQ)}
%%%%%%%%%%%%%%%%%%%%%%%%%%%%%%%%%%%%%%%%%%
In addition to usability, UX was assessed using the UEQ, which provides a differentiated evaluation across six dimensions on a scale ranging from $-3$ to $+3$, with higher values indicating a more positive UX. 
As commonly observed in UEQ-based evaluations, aggregated mean values typically fall within a relatively narrow range due to differences in individual response tendencies \citep{brooke2013sus}.
%%%%%%%%%%%%%%%%%%%%%%%%%%%%%%%%%%%%%%%%%%
\begin{figure}[!ht]
    \centering
    \includegraphics[width=\linewidth]{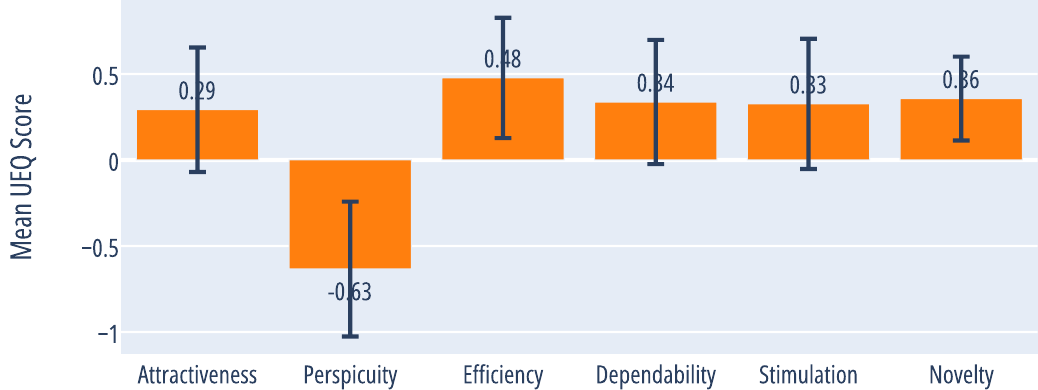}
    \caption{Mean scores and 95\% confidence intervals ($p = 0.05$) for the six User Experience Questionnaire (UEQ) dimensions of the Indicator Editor, measured on a scale from $-3$ (negative) to $+3$ (positive).}
    \label{fig:ueq-scales-mean-confidence}
\end{figure}
%%%%%%%%%%%%%%%%%%%%%%%%%%%%%%%%%%%%%%%%%%

Figure \ref{fig:ueq-scales-mean-confidence} presents the mean scores and confidence intervals ($p = 0.05$) for all UEQ dimensions\footnote{Adapted from the UEQ-Tool: https://www.ueq-online.org, last accessed on 2025-12-25}. 
Five of the six dimensions yielded positive mean values. 
Efficiency achieved the highest mean score ($0.48$), followed by dependability and stimulation, while attractiveness showed the lowest positive mean value ($0.29$). 
In contrast, perspicuity received a negative mean score of $-0.63$, indicating difficulties related to clarity, ease of understanding, and learnability.
The confidence intervals for all dimensions were relatively wide, suggesting substantial variability in participants' responses. 
This variability indicates heterogeneous UXs and limits the precision of the estimated mean values, making cautious interpretation of the results necessary.
The UEQ dimensions can be further grouped into higher-level quality constructs to support a more holistic interpretation of UX, namely attractiveness, pragmatic quality (perspicuity, efficiency, dependability), and hedonic quality (stimulation, novelty).
Pragmatic quality summarizes task-related aspects of interaction, reflecting how clearly, efficiently, and predictably users can accomplish their goals. 
Hedonic quality summarizes non-task-related aspects, reflecting users' motivation, engagement, and perceived innovativeness of the system.
%%%%%%%%%%%%%%%%%%%%%%%%%%%%%%%%%%%%%%%%%%
\begin{figure}[!ht]
    \centering
    \includegraphics[width=\linewidth]{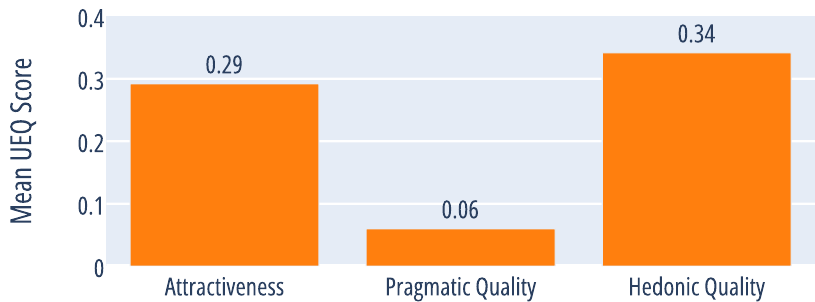}
    \caption{Aggregated mean UEQ scores for overall attractiveness, pragmatic quality (perspicuity, efficiency, dependability), and hedonic quality (stimulation, novelty) for the Indicator Editor.}
    \label{fig:ueq-pragmatic-hedonic}
\end{figure}
%%%%%%%%%%%%%%%%%%%%%%%%%%%%%%%%%%%%%%%%%%

As shown in Figure \ref{fig:ueq-pragmatic-hedonic}, pragmatic quality yielded an overall mean score of $0.06$, indicating a near-neutral evaluation of task-related interaction quality. 
This result is largely driven by the strongly negative perspicuity score observed in Figure \ref{fig:ueq-scales-mean-confidence}, which reflects confusion, difficulties in understanding system concepts, navigating the workflow, and learning how to use the tool. 
Although efficiency and dependability were rated more positively, these strengths were insufficient to compensate for the lack of clarity and understandability.
Hedonic quality also remained close to neutral, indicating that non-task-related aspects of the UX did not consistently support motivation and positive engagement. 
While participants perceived the system as stimulating and novel, these impressions were moderated by frustration caused by complexity, uncertainty, and limited predictability, which reduced the overall experiential appeal.
Taken together, the aggregated results indicate that both task-related qualities (e.g., clarity, learnability, and predictability) and non-task-related qualities (e.g., engagement and perceived novelty) require further improvement to achieve a consistently positive UX.
%%%%%%%%%%%%%%%%%%%%%%%%%%%%%%%%%%%%%%%%%%
\subsubsection{UEQ Dimension-Level Analysis}
\begin{figure*}[!t]
    \centering
    \includegraphics[width=0.75\textwidth]{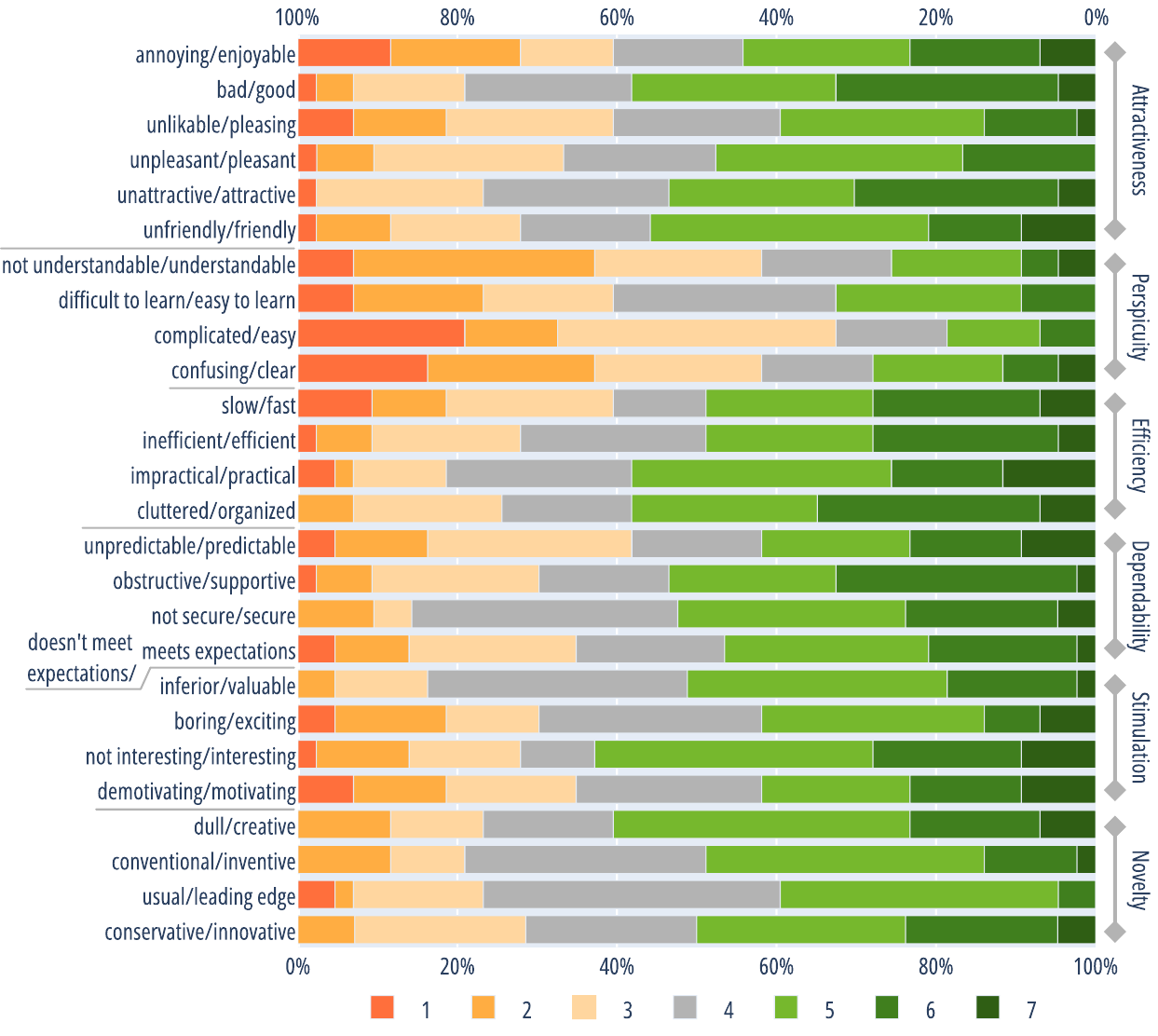}
    \caption{UEQ item-level response distributions for the Indicator Editor.}
    \label{fig:ueq-scores}
\end{figure*}

%%%%%%%%%%%%%%%%%%%%%%%%%%%%%%%%%%%%%%%%%%
To complement the aggregated UEQ scale scores, we further analyzed responses at the level of individual UEQ dimensions. 
For each dimension, we summarize the response distribution (Figure \ref{fig:ueq-scores}) and illustrate salient perceptions using representative excerpts from participants' open-ended responses collected at the end of the workshop. 
Quotes are anonymized and labeled as \textbf{P\#}, referring to individual workshop participants.
%%%%%%%%%%%%%%%%%%%%%%%%%%%%%%%%%%%%%%%%%%
%%%%%%%%%%%%%%%%%%%%%%%%%%%%%%%%%%%%%%%%%%

%%%%%%%%%%%%%%%%%%%%%%%%%%%%%%%%%%%%%%%%%%
\paragraph{Attractiveness}
%%%%%%%%%%%%%%%%%%%%%%%%%%%%%%%%%%%%%%%%%%
It describes the overall appeal of a product. 
It measures the user's general impression and how much they like or dislike the product. 
As shown in Figure \ref{fig:ueq-scores}, the participants have a positive impression of the \textit{Indicator Editor} in terms of attractiveness. 
Notably, they found its design to be good ($58\%$), friendly ($56\%$), and attractive ($53\%$), as indicated by one participant: \textbf{P11}: \textit{``Its design is clear and does not distract me from the main task.''}
Additionally, the variety of available visualizations enhanced the attractiveness of the \textit{Indicator Editor}, with another participant noting: \textbf{P14}: \textit{``I like how I can choose from various visualizations easily.''}
However, it can be observed that the disagreement was relatively higher regarding the \textit{Indicator Editor} being enjoyable, pleasing, and pleasant ($40\%$, $40\%$, $33\%$, respectively).
%%%%%%%%%%%%%%%%%%%%%%%%%%%%%%%%%%%%%%%%%%
\paragraph{Perspicuity}
%%%%%%%%%%%%%%%%%%%%%%%%%%%%%%%%%%%%%%%%%%
It evaluates how easily users can understand and become familiar with the product. 
It considers factors such as ease of learning, clarity, and how intuitive the product feels to the user.
As shown in Figure \ref{fig:ueq-scores}, the participants have a negative impression of the \textit{Indicator Editor} in terms of perspicuity on all four items. 
Particularly, with the abundance of options, the participants found the interface to be complicated ($67\%$) and difficult to learn ($40\%$), which overwhelmed them, \textbf{P16}: \textit{``Sometimes there are too many options to choose from, which can be overwhelming for someone new to the system''}, \textbf{P6}: \textit{``There are many new features that I don't know about.''} 
Additionally, users found the tool confusing ($58\%$), as one stated \textbf{P14}: \textit{``It gets confusing how my inputs are contributing to the changes in the next steps.''}
Furthermore, some participants found the placement and distinction between \textit{Activity Type} and \textit{Actions on Activities} under \textit{Dataset} section confusing, noting that these options did not intuitively belong there. 
Many participants suggested enabling multiple activity filters by allowing independent selection of \textit{Activity Types} and \textit{Action on Activities}. 
This approach may be perceived as a way to significantly reduce information overload and improve consistency within the \textit{Filters} section. 
Some options were not understandable ($58\%$), \textbf{P11}: \textit{``Some options are unclear, and it's hard to understand what the editor is asking me to do.''} 
Another user noted the difficulty in understanding some of the available data analysis methods, \textbf{P17}: \textit{``I do not know how to analyze data beyond the counting method.''}
Moreover, some participants indicated that the relationship between analytics method options and visualization options was one of the sources of confusion and lack of understandability, \textbf{P12}: \textit{``It's not clear which options [analysis methods outputs] correspond to which elements [visualization inputs] in the graph.''}
After receiving verbal explanations, participants suggested that providing a preview of the analyzed data before selecting the visualization could have helped them better understand this connection.
Moreover, some charts were deemed unsuitable because the outputs generated by specific analysis methods did not meet the input requirements of the selected visualizations. 
This mismatch confused participants and hindered them from selecting compatible visualizations.
%%%%%%%%%%%%%%%%%%%%%%%%%%%%%%%%%%%%%%%%%%
\paragraph{Efficiency}
%%%%%%%%%%%%%%%%%%%%%%%%%%%%%%%%%%%%%%%%%%
It measures how quickly and efficiently users can complete tasks using the product. 
It includes aspects such as the speed of task performance, the product's responsiveness, and how smoothly users can achieve their goals.
As shown in Figure \ref{fig:ueq-scores}, participants have mostly positive impressions of the \textit{Indicator Editor} in terms of efficiency.
Notably, they found the system to be organized and practical (both $58\%$), as mentioned by the participants: \textbf{P6}: \textit{``I can easily put my ideas into practice''}, \textbf{P13}: \textit{``You could see a preview of the visualization''}, \textbf{P15}: \textit{``Allowing us to make necessary changes.''}
However, many participants found the \textit{Indicator Editor} to be slow ($40\%$).
Moreover, relatively many participants found the \textit{Indicator Editor} to be inefficient ($28\%$) as they found the tool's complexity, multiple selections, and editing capability could hinder task completion, for example, \textbf{P11}: \textit{``Sometimes these [selections] hinder more than it helps, and the option to edit my indicators to fix some selections were not possible.''}
%%%%%%%%%%%%%%%%%%%%%%%%%%%%%%%%%%%%%%%%%%
\paragraph{Dependability}
%%%%%%%%%%%%%%%%%%%%%%%%%%%%%%%%%%%%%%%%%%
It assesses the user's sense of control and security while using the product. 
It covers reliability, predictability, and how many users feel they can trust the product to perform consistently and accurately. 
As shown in Figure \ref{fig:ueq-scores}, participants have an overall positive impression of the \textit{Indicator Editor} in terms of dependability.
They found the tool supportive ($53\%$). 
For example, participants appreciated its explanatory features, \textbf{P5}: \textit{``Usually provides explanations when needed''}, and valued the availability of pre-implemented analysis methods, \textbf{P28}: \textit{``I liked the fact that there were some already implemented analysis methods which were readily available to users.''}
They also found it secure ($52\%$) mainly due to the saving feature for created indicators, \textbf{P9}: \textit{``Includes a history of the indicators that I created.''}
However, concerns about predictability and meeting expectations were noted by relatively many participants ($42\%$ and $35\%$, respectively), for example, \textbf{P14}: \textit{``I can't control the outcomes.''}
This suggests more straightforward communication to enhance participants' sense of control.
%%%%%%%%%%%%%%%%%%%%%%%%%%%%%%%%%%%%%%%%%%
\paragraph{Stimulation}
%%%%%%%%%%%%%%%%%%%%%%%%%%%%%%%%%%%%%%%%%%
It refers to users' excitement and motivation towards the product. 
It reflects the product's ability to engage users, maintain their interest, and provide a stimulating experience.
As shown in Figure \ref{fig:ueq-scores}, the tool was perceived as interesting ($63\%$) and valuable ($51\%$), mainly due to its visual and interactive elements, for example, \textbf{P20}: \textit{``It shows a very easy visual of the data we have given while creating an indicator''} and \textbf{P23}: \textit{``I like that I am able to visualize my data immediately and edit as needed.''}
However, a relatively high number of participants ($35\%$) reported reduced motivation, mainly due to technical issues and system instability. For example, \textbf{P22} noted: \textit{``Many bugs. The site randomly shuts down and logs you out. It doesn't even save the results.''}
%%%%%%%%%%%%%%%%%%%%%%%%%%%%%%%%%%%%%%%%%%
\paragraph{Novelty}
%%%%%%%%%%%%%%%%%%%%%%%%%%%%%%%%%%%%%%%%%%
It measures how innovative and creative the product is perceived to be. 
It evaluates the product's uniqueness, originality, and how much it stands out compared to other similar products. 
As shown in Figure \ref{fig:ueq-scores}, participants predominantly perceived the \textit{Indicator Editor} as creative, innovative, and inventive ($60\%$, $50\%$, and $49\%$, respectively), indicating that many users perceived the system as distinct from existing analytics tools. In particular, users emphasized the novelty of systematically creating customized LA indicators based on real learning data, without programming skills. 
Nevertheless, participants perceived limitations that reduced the degree to which the system felt truly novel, mainly related to the lack of more visual customization of indicators after creation, as available in other visual analytics tools. 
%%%%%%%%%%%%%%%%%%%%%%%%%%%%%%%%%%%%%%%%%%
\subsubsection{Net Promoter Score (NPS)}
%%%%%%%%%%%%%%%%%%%%%%%%%%%%%%%%%%%%%%%%%%
To complement the usability and UX measures, we assessed participants' overall satisfaction and willingness to recommend the system using the NPS, measured on an 11-point scale from $0$ to $10$. 
Based on participants' responses after interacting with the \textit{Indicator Editor}, an overall NPS of $-28.3\%$ was obtained.
Specifically, $15.2\%$ of participants were classified as \textit{promoters} (ratings of $9$ or $10$), $41.3\%$ as \textit{passives} (ratings of $7$ or $8$), and $43.5\%$ as \textit{detractors} (ratings of $6$ or lower).
A negative NPS indicates that detractors outnumber promoters, suggesting limited willingness to actively recommend the system.
The contrast between the negative NPS and the acceptable SUS results suggests that, although participants were able to complete tasks successfully, their overall experience did not yet motivate strong advocacy or recommendation.

%%%%%%%%%%%%%%%%%%%%%%%%%%%%%%%%%%%%%%%%%%
\subsubsection{Synthesis of Findings}
%%%%%%%%%%%%%%%%%%%%%%%%%%%%%%%%%%%%%%%%%%
Taken together, SUS, UEQ, NPS, and qualitative feedback indicate a mixed evaluation profile of the \textit{Indicator Editor}, with clear strengths and weaknesses across task-related and non-task-related qualities.
\textit{Attractiveness} (good, friendly, attractive) and non-task-related qualities—\textit{stimulation} (interesting, valuable) and \textit{novelty} (creative, innovative)—were perceived positively, reflecting users' appreciation of the visual design, exploratory interaction, and the ability to create customized LA indicators without programming.
In contrast, task-related qualities revealed notable challenges. In particular, \textit{Perspicuity} (complicated, not understandable, confusing, difficult to learn) emerged as the most critical weakness, driven by option overload, unclear terminology, workflow mismatches, and difficulties relating analysis outputs to visualization inputs.
While \textit{efficiency} (organized, practical) was rated positively overall, interaction overhead and limited editability reduced task performance for some users. \textit{Dependability} showed mixed results: supportive features and persistence (secure) were valued, but concerns about predictability and user control remained (unpredictable, does not meet expectations).
The negative NPS suggests that these task-related issues outweighed positive impressions, limiting users' overall satisfaction and willingness to recommend the system.

%%%%%%%%%%%%%%%%%%%%%%%%%%%%%%%%%%%%%%%%%
\section{Final User Interface of the Indicator Editor} \label{final-system}
\begin{figure*}[!t]
	\centering
	\includegraphics[width=0.89\linewidth]{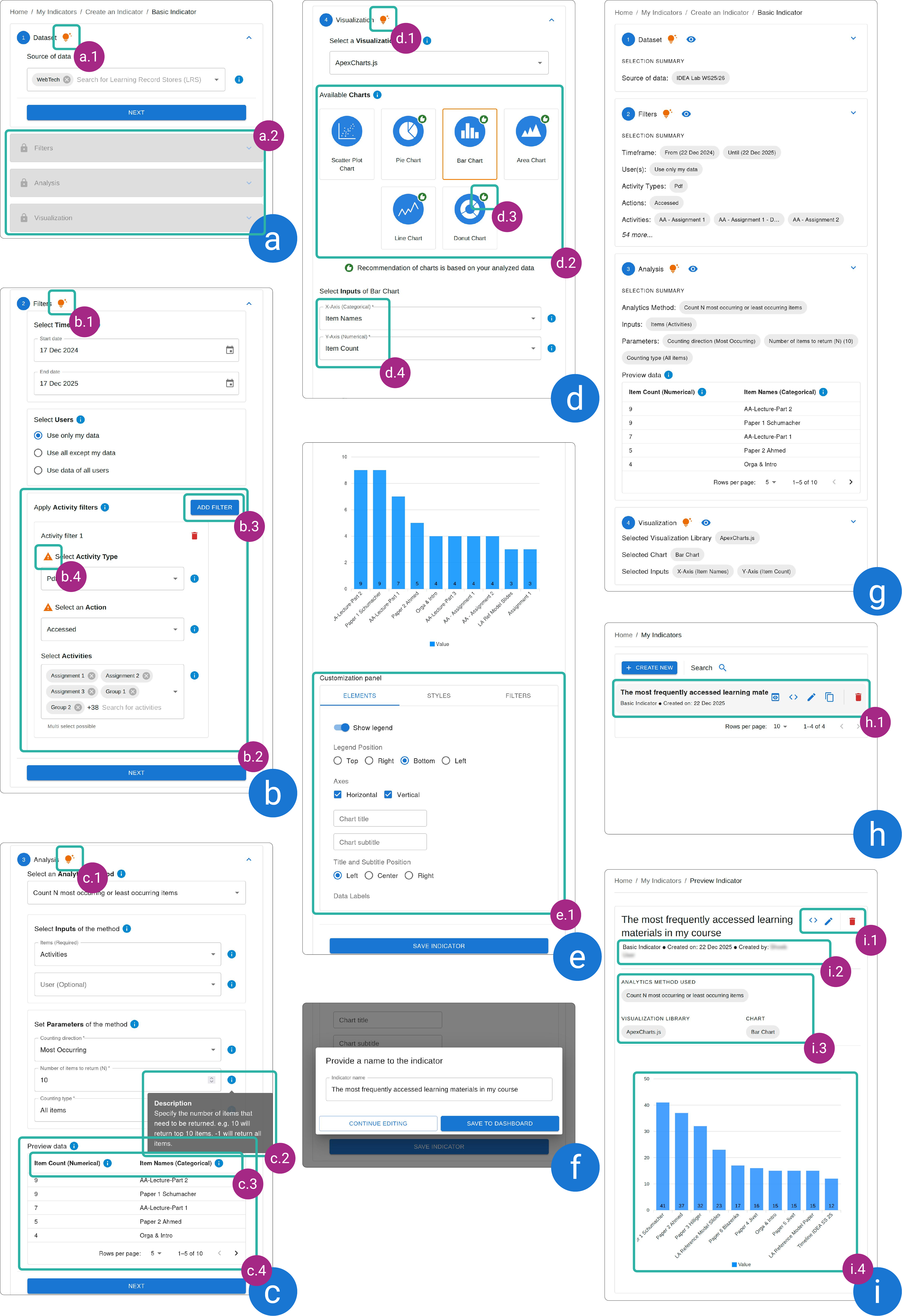}\\
    \caption{The Final User Interface of the Indicator Editor. The Editor screen supports dataset selection, filtering, analysis, visualization, preview/customization, saving, and review of a selection summary ($a$--$g$). The My Indicators screen supports indicator management and indicator preview ($h$--$i$).}
	\label{fig:final-user-interface}
\end{figure*}

%%%%%%%%%%%%%%%%%%%%%%%%%%%%%%%%%%%%%%%%%
Based on the usability and UX evaluation of the \textit{Intermediate High-Fidelity Prototype}, we refined the \textit{Indicator Editor} into its final user interface and deployed it to the Open Learning Analytics Platform (OpenLAP)\footnote{https://openlap.de/}. 
The redesign mainly targets key task-related weaknesses identified in the evaluation, specifically limited perspicuity (complicated, not understandable, confusing, difficult to learn), reduced efficiency, and dependability (unpredictable, does not meet expectations).
Figure \ref{fig:final-user-interface} presents the resulting \textit{Final User Interface of the Indicator Editor}, which comprises two main screens: 
(1) an \textit{Editor} screen that guides users through indicator implementation, and 
(2) a \textit{My Indicators} screen that supports revisiting, previewing, and managing created indicators.
%%%%%%%%%%%%%%%%%%%%%%%%%%%%%%%%%%%%%%%%%
%%%%%%%%%%%%%%%%%%%%%%%%%%%%%%%%%%%%%%%%%

\subsection{Editor Screen}
The Editor supports end-to-end indicator implementation through five configuration steps (dataset, filters, analysis, visualization, and finalization), as shown in Figure \ref{fig:final-user-interface}$a$--$g$.
To address perspicuity issues reported in the workshop (i.e., the process felt complicated and difficult to learn and users felt overwhelmed due to the abundance of options), the workflow is now presented as a vertically ordered set of collapsible sections with explicit step numbering and progress cues (\textbf{F1}) (Figure \ref{fig:final-user-interface}$a$). 
To reduce overload from too many simultaneously visible options, only the current step is emphasized, while later steps remain visually disabled until prerequisite selections are completed (\textbf{F3}) (Figure \ref{fig:final-user-interface}$a.2$). 
This progressive disclosure also improves dependability by making system behavior more predictable and reducing unexpected states (\textbf{F3}).
Across steps, the interface uses consistent section structure, control placement, and interaction behavior, addressing inconsistencies to increase predictability (dependability) and learnability (perspicuity) (\textbf{F4}).
To mitigate unclear terminology and ambiguous options, each step includes in-context help (\textbf{F2}) (Figure \ref{fig:final-user-interface}$a.1$, $b.1$, $c.1$, $d.1$). This would reduce confusion and enhance the understandability and learnability of the process (perspicuity).   

\subsubsection{Dataset}
The \textit{Dataset} section (Figure \ref{fig:final-user-interface}$a$) allows users to select the data source that forms the basis for all subsequent indicator configuration steps.
In contrast to earlier versions, selections related to \textit{Activity Type} and \textit{Actions on Activities} were removed from the \textit{Dataset} section and relocated to the \textit{Filters} section. 
This change responds to qualitative feedback and workshop observations that users perceived these options as conceptually mismatched with dataset selection. 
By separating data source selection from data refinement, the redesigned \textit{Dataset} section not only reduces confusion (perspicuity) but also aligns with users' mental models of the indicator implementation workflow (\textbf{F4}), thus improving conceptual consistency and predictability (dependability).

\subsubsection{Filters}
The \textit{Filters} section (Figure \ref{fig:final-user-interface}$b$) allows users to refine the selected dataset by specifying temporal constraints, user scope, and activity-related conditions. 
Users can define basic temporal and user-related filters directly, while additional activity-related filters are added explicitly via an \textit{Add Filter} button (Figure \ref{fig:final-user-interface}$b.3$).
This approach limits the number of simultaneously visible options (\textbf{F3}) and reduces interaction overhead by enabling incremental filter configuration.
Each \textit{Activity} filter is defined as a self-contained unit consisting of exactly one \textit{Activity Type}, one corresponding \textit{Action}, and one or more concrete \textit{Activities} (Figure \ref{fig:final-user-interface}$b.4$).
In contrast to earlier versions, which configured multiple \textit{Activity Types} and \textit{Actions on Activities} globally, this design allows users to add multiple independent activity filters, each capturing a single, coherent activity condition.
This restructuring directly addresses participants' confusion (perspicuity) regarding the placement and distinction of activity-related options and supports a clearer mental model by grouping logically related decisions within the \textit{Filters} section.
By constraining each filter to one activity type and one action, while still allowing multiple filters to be combined, the interface reduces information overload and makes filtering behavior more predictable (\textbf{F3}, \textbf{F4}).
Inline validation messages and warnings immediately indicate missing or incompatible selections, providing direct and actionable feedback (\textbf{F1}). 
These cues help users understand required inputs and reduce trial-and-error behavior.
To address terminology-related misunderstandings, filter options use plain-language labels and are supported by in-context explanations (\textbf{F2}). 

\subsubsection{Analysis}
The \textit{Analysis} section (Figure \ref{fig:final-user-interface}$c$) enables users to select an analysis method and configure how the filtered data are analyzed. 
Participants reported substantial difficulties understanding available analysis methods and their parameters, particularly users with limited data analytics experience. 
To improve understandability (perspicuity), available analysis methods and their parameters are described using plain, descriptive language rather than technical jargon (\textbf{F2}) (Figure \ref{fig:final-user-interface}$c.2$). 
Short in-context explanations clarify the purpose of each method and the meaning of configurable parameters, helping users select appropriate analyses without requiring prior expertise. 
To address confusion (perspicuity) about the relationship between analysis and visualization, the \textit{Analysis} section provides an intermediate representation of the processed data in the form of an immediate tabular preview (Figure \ref{fig:final-user-interface}$c.4$). 
This preview, together with metadata indicating numerical or categorical outputs (Figure \ref{fig:final-user-interface}$c.3$), allows users to inspect analysis results before visualization, reducing trial-and-error and improving predictability of subsequent choices (\textbf{F1}).
The separation of analysis method selection, parameter configuration, and data preview into clearly delineated subcomponents structures the task and limits cognitive load (\textbf{F3}). 

\subsubsection{Visualization}
The \textit{Visualization} section (Figure \ref{fig:final-user-interface}$d,e$) allows users to transform analyzed data into visual representations. 
Related to perspicuity, participants reported confusion and difficulties understanding which visualization types were suitable for a given analysis and how analysis outputs should be mapped to visualization inputs. 
To address data–chart incompatibilities, users first select a visualization library and are then presented only with chart types that are compatible with the structure of the analyzed data (Figure \ref{fig:final-user-interface}$d.2$). 
Recommended chart types are visually highlighted (Figure \ref{fig:final-user-interface}$d.3$), providing guidance while still allowing user choice. 
This recommendation reduces invalid selections, limits trial-and-error, and improves perspicuity while reducing information overload (\textbf{F3}).
Once a chart type is selected, users map analysis outputs to visual axes (Figure \ref{fig:final-user-interface}$d.4$). 
Input fields explicitly indicate whether numerical or categorical data are required, drawing on metadata generated in the \textit{Analysis} section. 
This makes data–chart dependencies explicit and addresses prior confusion (perspicuity) about how analysis outputs map to visualization inputs (\textbf{F1}).
A live preview of the visualization is displayed immediately after valid mappings are provided (Figure \ref{fig:final-user-interface}$e$). 
Any changes to mappings or visualization parameters are reflected instantly, reinforcing a clear cause--effect relationship between user actions and visual output and improving predictability (dependability) and feedback (\textbf{F1}). 
A unified customization panel allows users to personalize indicators (Figure \ref{fig:final-user-interface}$e.1$), thus supporting exploration and user control, and meeting user expectation (dependability). 

\subsubsection{Finalization and Selection}
After completing the visualization step, the tool presents a selection summary that consolidates all configuration decisions, including the selected dataset, applied filters, analysis method, and visualization settings (Figure \ref{fig:final-user-interface}$g$). 
This overview allows users to review and validate their choices holistically, addressing evaluation findings in which participants reported uncertainty about whether the system behaved as expected and whether outcomes could be reliably anticipated (\textbf{F1}).
Before saving the indicator, the users are prompted to name it (Figure \ref{fig:final-user-interface}$f$). 
This explicit finalization step marks the transition from configuration to completion and helps users recognize when the indicator implementation process is nearing completion, supporting orientation and feedback within the workflow (\textbf{F1}).

\subsection{My Indicators Screen}
The \textit{My Indicators} screen provides a centralized overview of all indicators created by the user and allows revisiting, editing, duplicating, and managing them. 
Participants reported limited possibilities to revise or correct indicators, which could hinder task completion and impact the tool's efficiency. 
The redesigned \textit{My Indicators} screen addresses this by making indicator management actions explicit, predictable, and easily accessible.
Each indicator is presented as a list entry containing its name and key metadata, such as indicator type and creation date (Figure \ref{fig:final-user-interface}$h.1$). 
This overview supports quick recognition and comparison, reducing uncertainty about what has been created (\textbf{F1}). 

The \textit{Preview} section (Figure \ref{fig:final-user-interface}$i$) provides a read-only, detailed view of an individual indicator before it is embedded in external dashboards or learning environments. 
At the top of the screen, consistently placed action buttons allow users to share indicator code, edit, or delete the indicator (Figure \ref{fig:final-user-interface}$i.1$), ensuring continuity with the interaction patterns established in the \textit{My Indicators} screen (\textbf{F4}).
Below this, users can find the indicator title and metadata, including creation date and creator (Figure \ref{fig:final-user-interface}$i.2$) (\textbf{F1}). 
The \textit{Preview} section also presents a structured summary of the analytical configuration, including the selected analysis method and visualization settings (Figure \ref{fig:final-user-interface}$i.3$). 
The resulting visualization is displayed prominently at the bottom of the screen (Figure \ref{fig:final-user-interface}$i.4$), allowing users to directly relate configuration decisions to the visual output. 
Separating metadata, configuration details, and visualization into clearly defined sections reduces cognitive load and supports focused inspection (\textbf{F3}).
%%%%%%%%%%%%%%%%%%%%%%%%%%%%%%%%%%%%%%%%%%
\section{Conclusion and Future Work} \label{conclusion}
%%%%%%%%%%%%%%%%%%%%%%%%%%%%%%%%%%%%%%%%%%
This paper presented a comprehensive usability evaluation and iterative improvement of the \textit{Indicator Editor}, a Self-Service Learning Analytics (SSLA) tool designed to support educational stakeholders in implementing custom learning analytics (LA) indicators. 
Motivated by the limited empirical investigation of usability aspects in prior evaluations of the \textit{Indicator Editor}, we combined an exploratory qualitative user study, usability inspections, and a workshop-based evaluation with students ($n=46$) using standardized instruments to examine how users interact with and perceive the \textit{Indicator Editor} in an authentic learning environment.
The results indicate that the \textit{Indicator Editor} achieves a good level of perceived usability and is experienced as attractive, friendly, practical, interesting, valuable, and novel. 
At the same time, the evaluation revealed recurring usability challenges related to perspicuity and predictability, particularly during complex configuration steps.
The implemented design improvements demonstrate how such challenges can be mitigated through clearer workflow communication, immediate and context-sensitive feedback, stepwise disclosure to reduce information overload, consistent interaction design, and plain-language explanations embedded directly within the interface.
Overall, this work contributes empirical insights and practical design guidance for researchers and practitioners aiming to develop usable and accessible SSLA tools that support effective indicator implementation.
Future work will extend the evaluation of the \textit{Final User Interface of the Indicator Editor} to additional stakeholder groups, such as teachers and learning designers, and investigate its long-term use in other authentic educational environments.

\bibliographystyle{apalike}
{\small\bibliography{example}}

@article{braun2006using,
  title     = {Using thematic analysis in psychology},
  author    = {Braun, Virginia and Clarke, Victoria},
  journal   = {Qualitative research in psychology},
  volume    = {3},
  number    = {2},
  pages     = {77--101},
  year      = {2006},
  publisher = {Taylor \& Francis}
}

@article{buckingham2019human,
  title   = {Human-centred learning analytics},
  author  = {Buckingham Shum, Simon and Ferguson, Rebecca and Martinez-Maldonado, Roberto},
  journal = {Journal of Learning Analytics},
  volume  = {6},
  number  = {2},
  pages   = {1--9},
  year    = {2019}
}

@incollection{chatti2021designing,
  title     = {Designing Theory-Driven Analytics-Enhanced Self-Regulated Learning Applications},
  author    = {Chatti, Mohamed Amine and Y{\"u}cepur, Volkan and Muslim, Arham and Guesmi, Mouadh and Joarder, Shoeb},
  booktitle = {Visualizations and Dashboards for Learning Analytics},
  pages     = {47--68},
  year      = {2021},
  publisher = {Springer}
}

@inproceedings{chatti2020design,
  title        = {How to design effective learning analytics indicators? A human-centered design approach},
  author       = {Chatti, Mohamed Amine and Muslim, Arham and Guesmi, Mouadh and Richtscheid, Florian and Nasimi, Dawood and Shahin, Amin and Damera, Ritesh},
  booktitle    = {European conference on technology enhanced learning},
  pages        = {303--317},
  year         = {2020},
  organization = {Springer}
}

@book{norman2013design,
  title     = {The design of everyday things: Revised and expanded edition},
  author    = {Norman, Don},
  year      = {2013},
  publisher = {Basic books}
}

@article{alfredo2024human,
  title={Human-centred learning analytics and AI in education: A systematic literature review},
  author={Alfredo, Riordan and Echeverria, Vanessa and Jin, Yueqiao and Yan, Lixiang and Swiecki, Zachari and Ga{\v{s}}evi{\'c}, Dragan and Martinez-Maldonado, Roberto},
  journal={Computers and Education: Artificial Intelligence},
  volume={6},
  pages={100215},
  year={2024},
  publisher={Elsevier}
}

@article{topali2025designing,
  title={Designing human-centered learning analytics and artificial intelligence in education solutions: a systematic literature review},
  author={Topali, Paraskevi and Ortega-Arranz, Alejandro and Rodr{\'\i}guez-Triana, Mar{\'\i}a Jes{\'u}s and Er, Erkan and Khalil, Mohammad and Ak{\c{c}}ap{\i}nar, G{\"o}khan},
  journal={Behaviour \& Information Technology},
  volume={44},
  number={5},
  pages={1071--1098},
  year={2025},
  publisher={Taylor \& Francis}
}

@misc{buckingham2024human,
  title={Human-centred learning analytics: 2019--24},
  author={Buckingham Shum, Simon and Mart{\'\i}nez-Maldonado, Roberto and Dimitriadis, Yannis and Santos, Patricia},
  journal={British Journal of Educational Technology},
  volume={55},
  number={3},
  pages={755--768},
  year={2024},
  publisher={Wiley Online Library}
}

@article{bangor2009determining,
  title={Determining what individual SUS scores mean: Adding an adjective rating scale},
  author={Bangor, Aaron and Kortum, Philip and Miller, James},
  journal={Journal of usability studies},
  volume={4},
  number={3},
  pages={114--123},
  year={2009},
  publisher={Usability Professionals' Association Bloomingdale, IL}
}

@inproceedings{laugwitz2008construction,
  title={Construction and evaluation of a user experience questionnaire},
  author={Laugwitz, Bettina and Held, Theo and Schrepp, Martin},
  booktitle={Symposium of the Austrian HCI and usability engineering group},
  pages={63--76},
  year={2008},
  organization={Springer}
}

@inproceedings{schrepp2014applying,
  title={Applying the user experience questionnaire (UEQ) in different evaluation scenarios},
  author={Schrepp, Martin and Hinderks, Andreas and Thomaschewski, J{\"o}rg},
  booktitle={International conference of design, user experience, and usability},
  pages={383--392},
  year={2014},
  organization={Springer}
}

@article{brooke2013sus,
  title={SUS: a retrospective.},
  author={Brooke, John},
  journal={Journal of usability studies},
  volume={8},
  number={2},
  year={2013}
}

@article{keiningham2008holistic,
  title={A holistic examination of Net Promoter},
  author={Keiningham, Timothy L and Aksoy, Lerzan and Cooil, Bruce and Andreassen, Tor Wallin and Williams, Luke},
  journal={Journal of Database Marketing \& Customer Strategy Management},
  volume={15},
  number={2},
  pages={79--90},
  year={2008},
  publisher={Springer}
}

@book{kuckartz2007einfuhrung,
  title={Einf{\"u}hrung in die computergest{\"u}tzte Analyse qualitativer Daten},
  author={Kuckartz, Udo},
  year={2007},
  publisher={Springer}
}

@book{dresing2007qualitative,
  title={Qualitative Evaluation: Der Einstieg in die Praxis},
  author={Dresing, Thorsten and Kuckartz, Udo and R{\"a}diker, Stefan},
  year={2007},
  publisher={VS Verlag f{\"u}r Sozialwissenschaften! GWV Fachverlage GmbH, Wiesbaden}
}

@article{joarder2025human,
  title={Human-Centred Development of Indicators for Self-Service Learning Analytics: A Transparency through Exploration Approach},
  author={Joarder, Shoeb and Chatti, Mohamed Amine},
  journal={Journal of Learning Analytics},
  year={2026},
  publisher={Society for Learning Analytics Research (SoLAR)}
}

@inproceedings{joarder2024anocode,
  title={A No-Code Environment for Implementing Human-Centered Learning Analytics Indicators},
  author={Joarder, Shoeb and Chatti, Mohamed Amine and Sun, Ao},
  booktitle={Companion Proceedings of the 14th International Learning Analytics and Knowledge Conference},
  pages={263--265},
  year={2024}
}

@article{ericsson2017protocol,
  title={Protocol analysis},
  author={Ericsson, K Anders},
  journal={A companion to cognitive science},
  pages={425--432},
  year={2017},
  publisher={Wiley Online Library}
}

@article{fan2019concurrent,
  title={Concurrent think-aloud verbalizations and usability problems},
  author={Fan, Mingming and Lin, Jinglan and Chung, Christina and Truong, Khai N},
  journal={ACM Transactions on Computer-Human Interaction (TOCHI)},
  volume={26},
  number={5},
  pages={1--35},
  year={2019},
  publisher={ACM New York, NY, USA}
}

@book{nielsen1994usability,
  title={Usability engineering},
  author={Nielsen, Jakob},
  year={1994},
  publisher={Morgan Kaufmann}
}

\end{document}